\documentstyle[12pt]{article}
\begin{document}

\begin{titlepage}
\begin{center}
{\Large {\bf Quark Delocalization, Color Screening\\
and Dibaryons\\}}

\vspace*{0.25in}

Fan Wang

Center for Theoretical Physics, Nanjing University\\
Nanjing, 210008, China\\

\vspace*{0.10in}

Jia-lun Ping

Department of Physics, Nanjing Normal University\\
Nanjing, 210024, China\\

\vspace*{0.10in}

Guanh-han Wu, Li-jiang Teng

Institute of Nuclear Science and Technology\\
Sichuan University, Chengdu,
610064, China\\

\vspace*{0.10in}

T. Goldman

Theoretical Division, Los Alamos National Laboratory\\
Los Alamos, NM87545,
USA
\end{center}

\begin{abstract}
The quark delocalization and color screening model, a quark potential\\
model, is used for a systematic search of dibaryon candidates in the
$u,d$ and $s$ three flavor world. Color screening which appears in
unquenched lattice gauge calculations and quark delocalization (which
is similar to electron delocalization in molecular physics) are both
included. Flavor symmetry breaking and channel coupling effects are
studied. The model is constrained not only by baryon ground state
properties but also by the $N$-$N$ scattering phase shifts. The
deuteron and zero energy di-nucleon resonance are both reproduced
qualitatively. The model predicts two extreme types of dibaryonic
systems: ``molecular'' like the deuteron, and highly delocalized
six-quark systems among which only a few narrow dibaryon resonances
occur in the $u,d$ and $s$ three flavor world. Possible high spin
dibaryon resonances are emphasized.
\end{abstract}

\end{titlepage}

\begin{center}
{\large {\bf I. INTRODUCTION}}
\end{center}

Quantum chromodynamics(QCD) is believed to be the fundamental theory of
the strong interaction. High energy processes are calculable due to the
asymptotic freedom property of QCD. The majority of low energy
processes are uncalculable due to infrared confinement. Lattice gauge
calculations may suffice in the confinement regime, but will still
suffer from large numerical uncertainties for the prediction of many
hadron properties in the near future. This leads to a reliance on QCD
inspired models to explore hadron physics for the time being and
perhaps even in the future, due to the complexity of QCD. The existing
models (potential, bag, soliton, etc.) are quite successful in
understanding hadron (meson and baryon) properties, but have not been
very successful for hadron interactions. Only recently have there been
positive indications for obtaining the whole $N$-$N$ interaction from
QCD models$^{[1,2]}$.

An outstanding problem is the fact that all of these models, including
lattice gauge calculations, predict that there should be: multiquark
systems $(q\bar q)^2,q^{4}\bar{q},q^6$; quark gluon hybrids
$q\bar{q}g,q^{3}g$; and glueballs in addition to the $q\bar{q}$ mesons
and $q^3$ baryons. Experimentally there are no well established
candidates for these exotics. In a relativistic theory, since quark and
gluon number is not conserved, any meson state can be a mixture of
$q\bar{q},(q\bar{q})^2,g^2$ and $q\bar{q}g$; any baryon state can be a
mixture of $q^3,q^{4}\bar{q}$ and $q^{3}g$. It is quite possible that
these exotics, $(q\bar{q})^2,q\bar{q}g,g^2,q^{4}\bar{q},q^{3}g$, exist
in the normal meson and baryon states$^{[3]}$. Polarized lepton
nucleon scattering measurements have aroused a new round of hadron
structure studies, wherein these exotic components are explored in
connection with the normal $q^3$ and $q\bar{q}$
components$^{[4]}$. However, $q^6$ is really a new quark system
sector, different from that of mesons and baryons. We call a baryon
number B=2 state, which is quasi-stable, a
dibaryon. Its minimum configuration is $q^6$.  Since Jaffe predicted
the first dibaryon, the $H$ particle$^{[5]}$, a large number of
dibaryon calculations with all the above-mentioned QCD models have been
carried out and almost all support the existence of
dibaryons$^{[6]}$.  If the present absence of an experimental
dibaryon signal continues, then all these QCD models (and even QCD
itself) should be questioned. Therefore, the dibaryon is a good place
to test QCD and its models.

Recently, Silvestre-Brac {\em et al.} reported a new systematic dibaryon
calculation based on the chromomagnetic model$^{[6]}$. As pointed out by
Lichtenberg and Roncaglia, the chromomagnetic model Hamiltonian is
oversimplified$^{[7]}$. The chromomagnetic interaction can give only the
$N$-$N$ short range repulsion but not any $N$-$N$ attraction. Many
dibaryon model calculations have the same deficiencies unless a
phenomenological meson exchange is invoked. To study the dibaryon, it
is better to have a model Hamiltonian which can at least fit the
$N$-$N$ interaction qualitatively. Then we can expect that such a model
prediction may be relevant to real dibaryon states. Another deficiency
of many prevailing model calculations is that the model Hilbert space
is rather restricted. In some model calculations$^{[5,6,8]}$, the six
quarks are assumed to be completely merged into a single confinement
region (which we term a `fully deconfined' model). In other model
calculations, the quarks are assumed to be always confined separately
in two distinct baryons (which we term a fully confined model)$^{[9]}$.
The real situation is quite possibly neither completely merged nor
always separately confined, but rather, in between, i.e., partially
deconfined due to the interaction of two baryons. A more realistic
model calculation allows the six quark system to choose the preferred
configuration by its own dynamics.

To remedy these model deficiencies, we developed a model which we will
call the quark delocalization, color screening model(QDCSM)$^{[2]}$. The
model Hilbert space is enlarged to include the fully confined and fully
deconfined models as two extremes and the real configuration is
determined variationally by the dynamics of the six quark system. In
this way the system is allowed to develop its own preferred distortion.
The model Hamiltonian is sufficiently realistic  to produce a
qualitatively correct $N$-$N$ phase shift. We especially take into
account the possible difference of the $q$-$q$ interaction inside a
hadron and between two colorless hadrons due to the nonlinearity of QCD
(see section II). We use this model to do a systematic search within
the $u,d,s$ quark three flavor world, expecting it to improve the
reliability of estimates on promising dibaryon candidates. This
expectation is realized in the $N$-$N$ channels (see section IV). Of
course we cannot expect that the model estimate is quantitatively
correct, because both the model Hamiltonian and Hilbert space are
restricted to be simple enough to do a systematic search. As emphasized
by Silvestre-Brac$^{[10]}$ this kind of systematic search  serves the
purpose of delimiting, among the thousands of multiquark states, the
most promising candidates. Our intent is to assist experimental efforts
to explore a challenging question in hadron physics by providing more
reliable theoretical estimates.

This paper is organized as follows. In section II, the model
Hamiltonian and Hilbert space are described. Section III is devoted to
a sketch of the calculation method. (A more complete description will
be reported separately). The results are given in section IV and a
conclusion in section V.
\vspace*{-0.10in}
\begin{center}
{\large {\bf II.\ QUARK DELOCALIZATION,\\
COLOR SCREENING MODEL}}
\end{center}

Quark potential models are quite successful in describing single
hadrons; therefore we adopt the usual potential model Hamiltonian to
describe a single baryon:
\begin{eqnarray}
H(3) & = & \sum_{i=1}^3 (m_i+\frac{p_i^2}{2m_i}) +\sum_{i<j=1}^{3}V_{ij}
       -T_c,  \\
V_{ij} & = & V_{ij}^c + V_{ij}^G, \nonumber \\
V_{ij}^c & = & -{\vec \lambda_i}\cdot {\vec \lambda_j}ar_{ij}^2, \nonumber \\
V_{ij}^G & = & \alpha_s \frac{ {\vec \lambda_i} \cdot {\vec \lambda_j} }{4}
 \left( \frac{1}{r_{ij}}-\frac{\pi}{2} \left( \frac{1}{m_i^2}+\frac{1}{m_j^2}
 +\frac{4}{3m_im_j} {\vec \sigma_i} \cdot {\vec \sigma_j} \right)
 \delta ({\vec r_{ij}) } + \cdots \right). \nonumber
\end{eqnarray}
The symbols in eq.(1) have their usual meaning. For the confinement
potential $V^c$, we assume a quadratic form to simplify the
calculation. A possible constant part is omitted to reduce the number
of parameters. In the effective one gluon exchange potential $V^G$,
only the color Coulomb and color magnetic terms are retained, because
we are only interested in the ground state. The effect of the momentum
dependent Darwin term has been checked and found not to be critical;
hence it is also omitted to reduce the calculational burden.

A variational three quark wave function (WF) of the form
\begin{equation}
\psi (123)=\chi (123)\eta _{SIJ}(123)\phi (123),
\end{equation}
is assumed to describe the ground state baryons. Here $\chi (123)$ is
the color singlet WF, $\eta _{SIJ}(123)$ is the symmetric spin-flavor
SU$_{2\times f}^{f\sigma }\ \supset $SU$_f\times $SU$_2^\sigma $ WF ($S
=$ strangeness, $I =$ isospin, $J =$ spin).
\renewcommand{\theequation}{3\alph{equation}} \setcounter{equation}{0}
\begin{equation}
\phi (123)=\phi _{1s}({\vec r_1})\phi _{1s}({\vec r_2})\phi _{1s}({\vec r_3}%
),
\end{equation}
\begin{equation}
\phi _{1s}({\vec r})= \left( \frac{1}{\pi b^2} \right)^{\frac{3}{4}}  e^{-
\frac{({\vec r}-{\vec s})^2}{2b^2}},
\end{equation}
${\vec s}$ is a reference center, and $b$ is a baryon size parameter to be
determined by the stability condition
\renewcommand{\theequation}{\arabic{equation}} \setcounter{equation}{3}
\begin{equation}
\frac{\partial M(123)}{\partial b} = 0,
\end{equation}

\noindent where $M(123) = \langle\psi (123)|H(123)|\psi (123)\rangle$.

The other model parameters are fixed as follows: the $u,d$ quark mass
difference is neglected and $m=m_u=m_d$ is assumed to be exactly $\frac
13$ of the nucleon mass M, {\em i.e}., $m=313\,$MeV. The quark gluon
coupling constant $\alpha _s$ is determined by the N-$\Delta $ mass
difference. The confinement potential strength $a$ is determined by the
zero nucleon binding. The strange quark mass $m_s$ is determined by an
overall fit to the strange baryon masses under the assumption that all
the flavor octet and decuplet baryons have the same r.m.s radius $b$.
(Choosing different values of $b$ for different baryons will make the
calculation more elaborate and is left as a future refinement.) The
fitted parameters are: $m=313\,$MeV, $m_s=634\,$MeV, $b=0.603\,$fm,
$\alpha_s$=1.54, $a=25.13\,$MeV$\cdot$fm$^{-2}$. The theoretical baryon
masses are compared with experimental values in Table I.

\begin{center}
Table I. Baryon masses (MeV)
\begin{tabular}{|l|c|c|c|c|c|c|c|c|} \hline
 & N & $\Delta$ & $\Lambda$ & $\Sigma$ & $\Xi$ & $\Sigma^*$ & $\Xi^*$ &
  $\Omega$ \\ \hline
exp.  & 939 & 1232 & 1115 & 1193 & 1318 & 1385 & 1533 & 1672 \\ \hline
theor.  & 939 & 1232 & 1118 & 1217 & 1359 & 1361 & 1504 & 1658 \\ \hline
\end{tabular}
\end{center}

The strange baryon masses do not agree perfectly due to our simple
model assumptions. The largest deviation is 41 MeV. The lower bounds on
the constituent quark mass differences derived from the Feynman-Hellman
theorem are fulfilled for our fitted quark masses, which are quite close
to those of Lichtenberg$^{[11]}$. The other parameters are
very similar in value to the usual baryon spectroscopic results$^{[12]}$.

The direct extension of the single baryon Hamiltonian eq.(1) to $q^6$
is neither reasonable nor successful. First, the two-body confinement
interaction will give rise to a spurious color van der Waals force. We
take this as an indication that the $q$-$q$ confinement interaction
between two color singlet hadrons is modified due to the nonlinearity
of QCD$^{[13]}$. The nonperturbative and lattice gauge approaches both
give rise to a string structure instead of a two-body $q$-$q$
interaction$^{[14]}$. The string structure and two-body confinement
give rise to similar spectroscopic results for simple quark system but
are not identical$^{[15,16]}$. Lattice gauge calculations, after taking
the (light) $q{\bar q}$ excitations into account, show that the (heavy)
$Q-{\bar Q}$ confinement interaction is screened. Numerical results can
be fitted by the following color screening interaction$^{[17]}$
\begin{equation}
V(r)=\left( -\frac{\alpha _s}r+\sigma r\right) \left( \frac{1-e^{-\mu r}}{%
\mu r}\right),
\end{equation}
$$
\alpha _s=0.21\pm 0.01,~~~ \sqrt{\sigma }=400\,\mbox{MeV}, ~~~ \mu
^{-1}=0.90\pm 0.20\,\mbox{fm}.
$$
Based on these results, we model the Hamiltonian of $q^6$ as an otherwise
direct extension of eq.(1), but modify the confinement part as
\renewcommand{\theequation}{6\Alph{equation}} \setcounter{equation}{0}
\begin{equation}
V_{ij}^c=\left\{
\begin{array}{ll}
-\overrightarrow{\lambda _i}\cdot {\vec \lambda}_j ar_{ij}^2 & ~~~
\mbox{if }i,j\mbox{ occur in the same baryon orbit}, \\
-{\vec \lambda}_i \cdot {\vec \lambda}_j \frac{a}{\mu}(1-e^{-\mu
r_{ij}^2}) & ~~~ \mbox{if }i,j\mbox{ occur in different baryon orbits}.
\end{array}
\right.
\end{equation}
Here the exponential $e^{-\mu r}$ appearing in eq.(5) has been replaced
by a Gaussian $e^{-\mu r^2}$, solely to simplify the numerical
calculations.  Another reason is this form will automatically match the
quadratic confinement in the short distance ($\mu r^2\ll 1$) region.
Keeping the same form of confinement as that of a single baryon when
the interacting pair of quarks occur in the same baryon guarantees that
when the two baryons are separated to large distances, the energy of
the $q^6$ system evolves to the two baryon internal energy calculated
by the Hamiltonian eq.(1).

B.Svetitsky$^{[18]}$ gave a qualitative description of the $Q-{\bar Q}$
potential: the short-range Coulomb potential evolves to a linear part
at larger distance, but at still larger distances, the linear part
evolves to an exponentially decaying Yukawa potential due to light meson
exchange. To take this possibility into account, we assume another
screening confinement potential (as described above).
\begin{equation}
V_{ij}^c=\left\{
\begin{array}{ll}
-{\vec \lambda}_i\cdot {\vec \lambda}_j ar_{ij}^2 & ~~~
\mbox{if }i,j\mbox{ occur in the same baryon orbit}, \\ -{\vec \lambda}_i
\cdot {\vec \lambda}_j ar_{ij}^2 e^{-\nu r_{ij}^2} & ~~~
\mbox{if }i,j\mbox{ occur in different baryon orbits}.
\end{array}
\right.
\end{equation}
Although eq.(6A) would be favored as better representing the Yukawa
fall-off to a constant potential at large distances, that constant is
not the zero value expected as between colorless hadrons. Thus, eq.(6B)
both tests the sensitivity (which turns out to be very mild) of our
model to the long distance behavior of the potential, and provides a
model with the expected zero value of the hadronic potential at infinity.

The screening parameter $\mu(\nu)$ is determined by fitting our model to the
N-N scattering phase shifts and they are$^{[2]}$
$$
\nu =0.40 \, \mbox{fm$^{-2}$},~~~~ \mu = 1.0\, \mbox{fm$^{-2}$}.
$$
Note particularly that the latter is consistent with the lattice
result$^{[17]}$.

For the $q^6$ model space, we extend the quark cluster model space by
introducing a delocalized single quark orbit. In the usual quark cluster
model approach, two single quark orbits are assumed
\renewcommand{\theequation}{\arabic{equation}} \setcounter{equation}{6}
\begin{equation}
\begin{array}{ll}
\phi_L({\vec r}) = \left(\frac{1}{\pi b^2} \right)^{\frac{3}{4}} e^{-\frac{({%
\ \vec r}+{\vec s}/2)^2}{2b^2}} ~~~~ & \mbox{(left centered orbit)}, \\
\phi_R({\vec r}) = \left(\frac{1}{\pi b^2} \right)^{\frac{3}{4}} e^{- \frac{(%
{\vec r}-{\vec s}/2)^2}{2b^2}} ~~~~ & \mbox{(right centered orbit)}.
\end{array}
\end{equation}
Here $s = \|{\vec s}\|$ is the separation of the centers of two $q^3$
cluster. We introduce the delocalized quark orbit,
\begin{eqnarray}
\psi_L({\vec r}) & = & \left( \phi_L({\vec r})+\epsilon(s) \phi_R({\vec r})
 \right)/N(s), \nonumber \\
\psi_R({\vec r}) & = & \left( \phi_R({\vec r})+\epsilon(s) \phi_L({\vec r})
 \right)/N(s), \\
N^2(s) & = & 1+\epsilon^2 (s)+2\epsilon (s)\langle\phi_L|\phi_R\rangle.
\nonumber
\end{eqnarray}
The delocalization parameter $\epsilon (s)$ is determined variationally
for every ${\vec s}$ by the dynamics of the $q^6$ system (see section
III). This orbit is a generalization of the quark molecular orbit
introduced by Fl.Stancu and L.Wilets$^{[19]}$, The six quark space is
restricted to be the space spanned by the following quark cluster
bases:
\begin{eqnarray}
\Psi^{\alpha}_{\alpha_1 F_1,\alpha_2 F_2} (1\cdots 6) & = & {\cal A}
  [\psi_{\alpha_1 F_1}(123)\psi_{\alpha_2 F_2}(456)]_{\alpha}, \nonumber \\
\psi_{\alpha_1 F_1}(123) & = & \chi (123) \eta_{S_1I_1J_1F_1}(123)\psi_L (1)
 \psi_L (2)\psi_L (3) \\
\psi_{\alpha_2 F_2}(456) & = & \chi (456) \eta_{S_2I_2J_2F_2}(456)\psi_R (4)
 \psi_R (5)\psi_R (6). \nonumber
\end{eqnarray}
Here $\alpha =(SIJ)$ describes the strong interaction conserved quantum
numbers of strangeness, isospin and spin.\footnote{Orbital angular
momentum is assumed to be zero for the lowest states. An angular
momentum projection which should be done is left for future work.
Preliminary estimates indicate this correction to the state energy is
small.} The $q^3$ cluster WF is almost the same as given in eq.(2), but
the single cluster Gaussian WF eq.(3b) is replaced by the delocalized
orbital WF eq.(8), and a flavor symmetry quantum number $F$ is shown
explicitly.  $[~]_\alpha $ refers to isospin and spin coupling by means
of the SU$_2$ Clebsch-Gordan coefficients. The SU$_3$ color coupling is
trivial because only color singlet hadron states are used in our
calculation. ${\cal A}$ is the normalized antisymmetry operator
$$
{\cal A}=\frac{1}{\sqrt{20} }\sum (-)^{\delta _p}p
$$
where $p$ is the two-quark permutation operator. Eq.(9) is termed the
physical basis by M.\ Harvey$^{[20]}$. Even though only totally
symmetric $q^3$ orbital configurations are included in eq.(9), $q^3$
orbital excitation configurations are included due to the delocalized
single quark orbit eq.(8) used in eq.(9). Hidden color channels are
{\em not} included in eq.(9). The reason is that the colorless channels
are already complete when orbital-spin-isospin excitation
configurations are included$^{[15,21]}$. Furthermore, color states have
not been well constrained in QCD models. It should be clearly kept in
mind that these physical bases are dependent on the separation ${\vec
s}$ of the two $q^3$ cluster centers and the delocalization parameter
$\epsilon (s)$.

A check had been done in the $N$-$N$ scattering dynamical calculation,
that if we start with the Fl.Stancu and L.Wilets molecular orbits
instead of the left and right centered orbits $\phi_L$ and $\phi_R$,
exactly the same results are obtained$^{[2]}$.

\begin{center}
{\large {\bf III. CALCULATION METHOD}}
\end{center}

A dynamical calculation of single channel $N$-$N$ scattering has been
done first to fix the screening parameter $\mu(\nu)$ by fitting the
model $N$-$N$ phase shifts to the experimental ones$^{[2]}$.

The $q^6$ states of a given set of quantum numbers $\alpha =(SIJ)$ are
expressed as a multiple physical channel coupling WF
\begin{equation}
\Psi _\alpha (1\cdots 6)=\sum_{\alpha _1F_1,\alpha _2F_2}C_{\alpha
_1F_1,\alpha _2F_2}^\alpha \Psi _{\alpha _1F_1,\alpha _2F_2}^\alpha .
\end{equation}
The channel coupling coefficients $C_{\alpha _1F_1,\alpha _2F_2}^\alpha
$ are determined by the diagonalization of the $q^6$ Hamiltonian. The
maximum number of channels coupled is 16. In the diagonalization, the
non-orthogonality property of the physical bases is properly
accounted for.

The six quark Hamiltonian matrix elements
$$
\langle\Psi_{\alpha^{\prime}_1 F^{\prime}_1,\alpha^{\prime}_2
F^{\prime}_2}^{\alpha}|H(1\cdots 6)| \Psi_{\alpha_1 F_1,\alpha_2
F_2}^{\alpha}\rangle
$$
are calculated by the group theory method developed by M.\ Harvey and
by J.\ Q.\ Chen and ourselves$^{[20]}$:

(1) The physical bases $\Psi _{\alpha _1F_1,\alpha _2F_2}^\alpha$ are
expressed in terms of the symmetry bases (group chain classification
bases) by the $6\rightarrow 3\times 3$, SU$_{mn}\supset$
SU$_m\times$~SU$_n$ and SU$_3^f\supset$ SU$_2^\tau \times $U$_1^S$
isoscalar factors calculated by Chen {\em et al\/}$^{[22]}$.

(2) The six quark Hamiltonian matrix elements in the symmetry bases are
reduced to a two body matrix element and a four quark overlap (due to
the non-orthogonality of the delocalized orbit) by the traditional
parentage expansion($6\rightarrow 4\times 2$, SU$_{mn}\supset$
SU$_m\times $SU$_n$ and SU$_3^f\supset$ SU$_2^\tau \times $U$_1^S$
isoscalar factor$^{[22]}$).

(3) The four quark overlap is reduced to a one body overlap by the
permutation symmetry property of the four quark state$^{[23]}$.

(4) Two body confinement interaction matrix elements are calculated as
follows: if the interacting quark pair occurs in the same left or right
orbit, {\em i.e}., $\langle LL|V|LL\rangle$, $\langle RR|V|RR\rangle$,
$\langle LL|V|RR\rangle$ and $\langle RR|V|LL\rangle$, then the usual
quadratic confinement form is used. We do so since the two quarks
involved are then always in the same `baryon', and hence in a relative
color anti-triplet state.  For all the other two body confinement
interaction matrix elements, such as $\langle LR|V|LR\rangle$,
$\langle~LL|V|LR\rangle$, $\cdots $, the color screening confinement
form is used.  Again, we do so since the confining interaction will
cancel over all the quark pairs identifiable as originating in
different color singlet hadrons; we effectively remove it in advance
from each pairwise interaction leaving an interaction which mimics
color singlet (mesonic) exchanges.  Here $\langle LL|V|RR\rangle$ means
$\langle\phi_L(1)\phi_L(2)|V_{12}|\phi_R(1)\phi_R(2)\rangle$.

The eigenenergies obtained in this way are dependent on the separation
${\vec s}$ and the delocalization parameter $\epsilon (s)$. We repeat
the calculation for each ${\vec s}$ by varying $\epsilon (s)$ from
0.1-1.0 with step size 0.1 to get a minimum of the eigenenergy, which
thus also determines the delocalization parameter $\epsilon (s)$.  The
difference of the minimum eigenenergy at separation ${\vec s}$ and the
minimum eigenenergy at infinite separation is taken to be the
baryon-baryon potential energy $V_\alpha (s)$ (an adiabatic
approximation). Numerically the asymptotic values of the eigenenergy
are indistinguishable from the calculation at $s=3\,$fm. It is equal to
the threshold sum, i.e., the sum of theoretical masses of the
corresponding channel baryon pair (single channel case) or of the
lightest baryon pair (channel coupling case), and the model relative
kinetic energy of this baryon pair, which is equal to $1/6$ of the
total kinetic energy of the $q^6$ system due to our model WF
assumption, eq.(9). (This is one of the checks of our numerical
calculation; another check is that all the channel mixing matrix
elements are vanishingly small at $s=3\,$fm).

A zero point harmonic oscillation energy $\frac{3\hbar^2}{4\mu s_0^2}$
is added to the minimum potential energy $V_{\alpha}(s_0)$ to obtain
the binding energy $B_{\alpha}$ of a $q^6$ state with the quantum number
$\alpha$.
\begin{equation}
B_{\alpha} = V_{\alpha}(s_0) +\frac{3\hbar^2}{4\mu_{\alpha}s_0^2}
\end{equation}
Here $\mu_{\alpha}$ is taken simply to be the reduced mass of the
corresponding channel baryon pair (single channel case) or the lightest
baryon pair within the quantum number $\alpha$ set (channel coupling case).
In principle, we should do a multichannel coupling dynamical calculation;
this program is being pursued only for the few most promising dibaryon
candidates.

Finally the experimental channel baryon pair mass (single channel case)
or the lightest baryon pair mass $(M_1+M_2)_\alpha $ (channel coupling
case) is added to $B_\alpha $ to obtain the lowest $q^6$ mass of each
quantum number $\alpha $ set,

\begin{equation}
M_\alpha (q^6)=(M_1+M_2)_\alpha +B_\alpha .
\end{equation}
The mass $M_\alpha (q^6)$ is compared not only to the two body decay
threshold, the experimental lightest baryon pair mass $(M_1+M_2)_\alpha $,
but also to the possible multi-particle final states allowed by strong
interaction to determine if there is a strong interaction quasi-stable
dibaryon state.

In order to show the effects of channel coupling, flavor-symmetry breaking

and the different forms of color screening, the following eight sets of
calculations have been done:

(1) single channel, flavor symmetry (scs),

(2) multi-channel, flavor symmetry (ccs),

(3) single channel, flavor symmetry breaking (scb),

(4) multi-channel, flavor symmetry breaking (ccb),

\noindent where (1)-(4) have been calculated using the color screening
form (6A) and (5)-(8) replace this with the color screening form (6B).

\noindent To indicate the level of uncertainty due to the choice of the
color screening parameter value, we have calculated the results
corresponding to $\mu = 1.6 fm^{-2}$ and $\nu = 0.46, 0.60 fm^{-2}$ in
addition to the best fit values, $\mu = 1.0 fm^{-2}$ and $\nu = 0.40
fm^{-2}$. (The reasoning behind the parameter values chosen for these
additional cases is described in the Appendix.) Only dibaryon
candidates based on both the channel coupling and symmetry breaking
results should be considered serious possibilities for experimental
searches.

A computer program package which incorporates all the needed group
theory results has been written to automate the numerical calculation.
It can be used for other model calculations by simply replacing the
one- and two-body matrix elements. In particular, it is may also be
used for a relativistic quark cluster model calculation. As a
cross-check on the program, the $S=0, (IJ)=(01),(10)$ and $(03)$
channels have been done both by the method described above and by
direct diagonalization with the physical bases.

\begin{center}
{\large {\bf IV. RESULTS}}
\end{center}

All possible sets of $\alpha=(SIJ)$ within the $u,d$, and $s$ three
flavor world have been calculated. Only a few states are strong
interaction quasi-stable or narrow resonances, and they are listed
in Table II.

The strong interaction unstable states have been omitted to simplify the
presentation. However some general features are listed here.

(1) The two color screening forms given quite similar results: The form
(6A) gives slightly higher six quark masses than the form (6B) but the
largest difference is only about 10$ \, Me$V for the $(SIJ)=(003)$
case. This is consistent with the findings for $N$-$N$ scattering where
the forms (6A) and (6B) give similar $N$-$N$ phase shifts while form
(6B) yields a little stronger attraction in the $^3S_1$
channel$^{[2]}$.

(2) There are two extreme kinds of dibaryon candidates. One kind is a
loosely bound (`molecular') two baryon state. The binding energy
$B_{\alpha}$ is small (usually around zero), $J\leq 1$; the
delocalization $\epsilon(s_0)$ is also small (usually
$\stackrel{<}{\sim} 0.2$). (The deuteron is a typical example while the
$H$ particle is an exception.) Their masses are close to the lowest two
body decay threshold and they might be stable with respect to the
strong interactions, but their stability is very sensitive to model
details. The other kind is a tightly bound state. The binding energy is
large ($\stackrel{>}{\sim} 100\,$MeV), $J\geq 2$; the delocalization is
also large ($\stackrel{>}{\sim} 0.8$). Their masses are larger than the
lowest two body threshold and they are therefore unstable. However
their masses are lower than the favorable three or four body decay
threshold. Their two or three body decay is hindered due to large
angular momentum and so they might nonetheless appear as narrow
resonances.  A typical example is the di-$\Delta$(003) state.  This
quasi-stability property is {\em not} sensitive to the model details,
unless radically different model assumptions are made.

(3) Flavor symmetry breaking effects are channel dependent as reported
by Malt-\\
man$^{[24]}$, decreasing the binding by an amount ranging from
nil to 70$\,$MeV. For some multi-channel coupling cases, the flavor
symmetry breaking effect changes which is the lowest channel.  In those
cases, there is large apparent flavor symmetry breaking effect ($\Delta
B_\alpha \sim 120-180$MeV). The mass of the $q^6$ state is increased
correspondingly and this produces a large difference in the stability
of the $q^6$ state with respect to the multi-particle decay channels,
due to the flavor symmetry breaking.

(4) The channel coupling effect is small for most cases, even after
taking into account quark delocalization in the extended $u,d,s$ world.
For the $H$ particle case, channel crossing occurs: The dominant
channel for the minimum potential ($\Sigma \Sigma$) and the asymptotic
channel ($\Lambda\Lambda$) are different. This channel coupling
decreases the $q^6$ mass by about 100 $\,$MeV and makes the $H$ particle
stable. A similar channel coupling effect occurs in the ($SIJ$)=(-202)
state.

We emphasize that all the QDCSM parameters are fixed by the ground
state baryon properties and the $N$-$N$ scattering phase shifts.
Therefore the dibaryon candidates noted here are a relatively robust
theoretical prediction. We wish to emphasize a few additional points.

(1) In the $S=0,(IJ)=(01)$ channel (the deuteron channel), the model
predicts that there is a state with $M(001)=1880 (\pm 14)\,$MeV. It is a
di-nucleon state because the delocalization $\epsilon = 0.2 $ is
small and 4 $\,$MeV away from the deuteron energy. We take this as a
measure of the predictive power of QDCSM for the dibaryon state. Due to
this uncertainty, we have included those states which are close to the
lowest two body decay threshold in table II.

(2) In the $S=0,(IJ)=(10)$ channel, the model predicts that there is a
di-nucleon ($\epsilon (s_0)=0.1$) resonance state at $1889 (\pm 4)
\,$MeV. It is about 10 MeV away from a possible zero binding di-nucleon
resonance.  This is another example that shows the predictive power of
QDCSM and appears to limit the model uncertainty.

(3) In the nonstrange sector ($S=0$), the model predicts an $\alpha
=(003)$ state with mass $M(003)=2110 (\pm 36) \,$MeV. It has the largest
binding (320-390 $\,$MeV) within the $u,d,s$ three flavor world.
Although its mass is above  $NN\pi$ threshold, the transition to
$NN\pi$ and $NN$ is hindered by the large angular momentum, so that it
is still possible for this to be a narrow resonance. The large
delocalization, $\epsilon =1.0$, means this is a true six quark state.
All these results are consistent with our earlier simple relativistic
model result$^{[25]}$.  Although the skyrmion model calculation of Walet
does not obtain a binding as large as QDCSM result$^{[26]}$, we note that
this model doesn't obtain sufficient attraction in the $N$-$N$ channel,
either. We know of no reason for the predictive power of the QDCSM
shown in the $N$-$N$ channel to be totally lost in the
$\Delta$-$\Delta$ channel since both channel results are stable under
the same reasonable variations in the values of the model parameters.
Therefore, we continue to recommend this state highly as a good
candidate for discovery of a dibaryon resonance.

There is no other interesting channel in the $u,d$ two flavor world as
found earlier by Maltman$^{[9]}$.

(4) Jaffe's $H$ particle remains as the unique strong interaction
stable dibaryon in QDCSM. $M(H)=2199 (\pm 24) \,$MeV is 32 $\,$MeV
lower than the $\Lambda \Lambda $ threshold. The delocalization
$\epsilon (s_0)=1.0$ is also large and the adiabatic channel coupling
WF is quite close to Jaffe's pure symmetric flavor singlet basis.
However, due to the sensitivity to details of the model, it is not
possible to claim that it is indeed a strong interaction stable state.

(5) Another interesting state is the $\alpha =(-3\frac 122)$
state$^{[27]}$.  $M(-3\frac 122)=2529 (\pm 25) \,$ MeV is 45$\,$MeV
lower than the favorable ($\Lambda \Xi \pi $) three body decay
threshold.  Also due to large angular momentum, its decay into $\Lambda
\Xi $ should be inhibited and so it too might show up as a dibaryon
resonance.  Another good point about this state is that all the other
states with the same quantum number set are about 100$\,$MeV higher
than it (not listed in table II). This might make it a cleaner
resonance to observe.

(6) The states $M(-1\frac{3}{2}0) = 2133(\pm 7)\,$MeV,$M(-220) =
2390(\pm 9)\,$ MeV,$M(-3\frac{3}{2}1)$\\ $= 2512(\pm 15)\,$ MeV,
$M(-3\frac{1}{2}1) = 2469(\pm 58)\,$ MeV, $M(-400) = \; 2637 (\pm
10)\,$ MeV,$M$\\ $(-600) = 3359(\pm 18)\,$ MeV, all have their masses
close to the corresponding thresholds: N$\Sigma$, $\Sigma\Sigma$,
$\Sigma\Xi$, $\Lambda\Xi$, $\Xi\Xi$ and $\Omega\Omega$. Another group
of states $M(-1\frac{1}{2}3)=2318(\pm 33)$\\ MeV, $M(-213)=2530(\pm
37)\,$MeV, $M(-202)=2345(\pm 26)\,$MeV, $M(-3\frac{3}{2}3) = 2728\\
(\pm 37)\,$MeV, have very large binding and their masses are all less
than the favorable multi-body channel. These states bear further study.
The high spin dibaryon resonances seem to be especially worth
experimental searches, in addition to the spin zero $H$ particle.

\begin{center}
{\large {\bf V. CONCLUSION}}
\end{center}

Since Jaffe's first prediction of H particle$^{[5]}$, there have been
many efforts both theoretically and experimentally to search for
dibaryons.  Whether all these QCD inspired models miss some physics
when they are extended from the single hadron to the multi-hadron case,
so that their predictions are not reliable, remains a question. The
QDCSM is a more realistic model by taking into account the possible
difference of the $q-q$ interaction inside a single baryon and between
two color singlet baryons, by allowing each system to choose its
favorable configuration in a larger Hilbert space, and by having the
model constrained not only by qualitatively fitting hadron spectroscopy
but also $N$-$N$ scattering.  This model approach has some moderate
success to support it: It predicts two di-nucleon states not too far
from the experimentally known deuteron and quasi-deuteron states. If we
take this as a measure of the predictive accuracy of the QDCSM, then
there are few promising dibaryon resonance candidates within the $u,d$
and $s$ three flavor world, as listed in table II. Due to the
simplicity of the model assumptions, the quantitative predictions of
the dibaryon masses are uncertain ($\sim 10\,$MeV for nonstrange states
and even larger uncertainty for strange states). To get more reliable
estimates, especially to be able to determine whether or not the
candidate states are strong interaction stable, requires improvement of
many aspects of the QDCSM:

(1) The QDCSM does not fit the baryon octet and decuplet perfectly, the
largest deviation being 41 MeV for the $\Xi $. Although the adiabatic
potential is obtained through a subtraction procedure which suggests
cancellation of errors is possible, there is no guarantee that the
uncertainty of the strange baryon mass cancels very accurately.

(2) An adiabatic approximation has been used in this calculation which
should be replaced by a dynamical channel coupling calculation.

(3) This calculation is nonrelativistic; a relativistic calculation is
underway to estimate relativistic corrections. The preliminary result
is that the relativistic and nonrelativistic versions give very similar
mass values, especially for the nonstrange states.

(4) Only $N$-$N$ scattering has been used to constrain the QDCSM.
Although data is sparser, $\Lambda$-p and $\Sigma$-p scattering should
be used as well. We have begun such an analysis.

(5) The effects of $q{\bar q}$ excitations or quark-meson couplings,
which have not been included, may well be important, especially for
those states which have a mass close to the lowest two-body threshold.

(6) It would be interesting to include $c$ and $b$ quarks with a view
towards making contact with Heavy Quark Effective Theory. However, the
large quark WF difference between $u,d$ and $b,c$ would need to be
treated first. This is unlike the $s$ quark case where the single quark
WF distortion is not large.

We believe the QDCSM results support the value of investing additional
effort, both theoretical and experimental, using more sophisticated
approaches, in order to concentrate on a few promising dibaryon
candidates.

\begin{center}
{\large {\bf ACKNOWLEDGMENTS}}
\end{center}

This work is supported by NSFC, the fundamental research project of the
state Science and Technology Committee, the graduate study fund of the
State Education Committee of China and the US DOE.

\pagebreak

\begin{center}
{\large {\bf APPENDIX}}
\end{center}

In the dynamical N-N scattering calculation$^{[2]}$, we first calculate the
interaction kernel
$$
K (\stackrel{\rightharpoonup}{S}, \stackrel{\rightharpoonup}{S}) =
\frac{< \Psi (\stackrel{\rightharpoonup}{S}) | H | \Psi
(\stackrel{\rightharpoonup}{S}) >}{< \Psi
(\stackrel{\rightharpoonup}{S}) | \Psi (\stackrel{\rightharpoonup}{S})
>} \eqno (A1)
$$
where $H$ is the six quark Hamiltonian, $\Psi
(\stackrel{\rightharpoonup}{S})$ is the N-N channel WF (9).  Then we do
a partial wave decomposition
$$
k_l (S, S^{\prime}) = \int d \Omega P_l (\ominus_{s s^{\prime}}) K
(\stackrel{\rightharpoonup}{S}, \stackrel{\rightharpoonup}{S^{\prime}})
\eqno (A2)
$$
where $\ominus_{s s^{\prime}}$ means the angle between
$\stackrel{\rightharpoonup}{S}$ and
$\stackrel{\rightharpoonup}{S^{\prime}}$.  We then assume the diagonal
matrix elements $k_l (s, s)$ as the effective interaction of the
l-partial wave between two nucleons,
$$
V_l (s) = k_l (s, s) \eqno (A3)
$$
This effective interaction is dependent on the delocalization parameter
$\epsilon (s)$.  We vary the value $\epsilon (s)$ to get a minimum for
each separation $s$.  The $\epsilon(s)$ so determined is $l$ dependent,
i.e., we have $\epsilon_l (s)$.  Next we substitute the values
$\epsilon_l (s)$ and $\epsilon_l (s^{\prime})$ back into eq.\ (A2) to
get the final $k_l (s, s^{\prime})$ for the N-N scattering
calculation.  Finally, we adjust the screening parameter $\mu$ or $\nu$
of eq.\ (6A) and (6b) to get the best fit to the $^1S_0$ and $^3S_1$
N-N phase shifts.  This determines the best values:  $\mu = 1.0
fm^{-2}$, $\nu = 0.4 fm^{-2}$, because it is a more complete and
consistent calculation.  However it takes much more computer time than
the next calculation we describe.

To minimize computer time for systematic dibaryon search, we tried
another approximation, assuming $K (\stackrel{\rightharpoonup}{s},
\stackrel{\rightharpoonup}{s})$ as the effective interaction between
two nucleons
$$
V (s) = K ( \stackrel{\rightharpoonup}{s}, \stackrel{\rightharpoonup}{s} )
\eqno (A4)
$$
Then, we varied $\epsilon (s)$ to minimize $V (s)$ and so determine the
$\epsilon (s)$ as well.  This $\epsilon (s)$ is partial wave
independent.  Substituting the $\epsilon (s)$ and $\epsilon
(s^{\prime})$ values so determined back to eq. (A1), we obtain the
final $K (\stackrel{\rightharpoonup}{s},
\stackrel{\rightharpoonup}{s^{\prime}})$ and then do a partial wave
decomposition to calculate the phase shifts.  Adjusting the screening
parameters again to obtain the best fit, the $^1 S_0$ and $^3 S_1$
channels, we obtain a second set of values:  $\mu = 1.6 fm^{-2}$, $\nu
= 0.6 fm^{-2}$.  The approximation is not as good as the first one, but
also gives a qualitatively good fit to the N-N phase shifts.

In the dibaryon calculation, we use the second variation method
(variation before partial wave decomposition) to obtain the effective
interaction between two baryons.  The results are shown in Tables IIa
and IIc.  As a check on the range of variation, we also used the
intermediate value $\nu = 0.46 fm^{-2}$, and those results are shown in
Table IIb.

\pagebreak

\pagebreak

\vspace*{-1.20in}
\begin{center}
Table IIa.  $\mu = 1.0 fm^{-2}, \nu = 0.4 fm^{-2}$\\
{\footnotesize
\begin{tabular}{|c|c|cllcc||cllcc|l|} \hline
   &   & \multicolumn{5}{c||}{(6B)} &  \multicolumn{5}{c|}{(6A)} & \\ \hline
$SIJ$ &  & M$_{\alpha}$ & $V_{\alpha}$ & $B_{\alpha}$ & $\epsilon$ & $s_0$ &
 M$_{\alpha}$ & $V_{\alpha}$ & $B_{\alpha}$ & $\epsilon$ & $s_0$ & Threshold
  \\ \hline
001 & scs & 1885 & -21 & 7 & 0.1 & 1.5 & 1885 & -20 & 7 & 0.1 & 1.5 &
  1878(NN) \\
    & ccs & 1886 & -23 & 8 & 0.2 & 1.4 & 1894 & -21 & 16 & 0.2 & 1.3 & \\
010 & scs & 1891 & -11 & 13 & 0.1 & 1.6 & 1893 & -10 & 15 & 0.1 & 1.6 &
  1878(NN) \\
    & ccs & 1891 & -11 & 13 & 0.1 & 1.6 & 1892 & -9 & 14 & 0.1 & 1.6 & \\
003 & scs & 2134 & -363 & -330 & 1.0 & 1.2 & 2144 & -359 & -320 & 1.0 & 1.1 &
    2464($\Delta\Delta$) \\
    &   &  & &  &  &  & &  &  &  &  & 2158(NN$\pi\pi$)  \\ \hline
-1$\frac{1}{2}$3 & scs & 2285 & -363 & -332 & 1.0 & 1.2 & 2294 & -359 & -322
 & 1.0 & 1.1 & 2617($\Delta\Sigma^*$) \\
     & ccs & 2285 & -363 & -363 & 1.0 & 1.2 & 2294 & -359 & -322 &
  1.0 & 1.1 & 2335(N$\Lambda\pi\pi$) \\
    & scb & 2343 & -311 & -274 & 1.0 & 1.1 & 2346 & -308 & -271 & 1.0 & 1.1 &
    \\
    & ccb & 2343 & -311 & -274 & 1.0 & 1.1 & 2346 & -308 & -271 & 1.0 & 1.1 &
\\
1$\frac{3}{2}$0 & scs & 2133 & -24 & 1 & 0.1 & 1.5 & 2134 & -23 & 2 &
  0.1 & 1.5 & 2132(N$\Sigma$) \\
    & ccs & 2133 & -24 & 1 & 0.1 & 1.5 & 2134 & -23 & 2 &
  0.1 & 1.5 &  \\
    & scb & 2137 & -19 & 5 & 0.1 & 1.5 & 2138 & -18 & 6 &
  0.1 & 1.5 &  \\
    & ccb & 2137 & -19 & 5 & 0.1 & 1.5 & 2138 & -18 & 6 &
  0.1 & 1.5 &  \\
  \hline
-200 & scs & 2145 & -195 & -112 & 0.5 & 0.8 & 2143 & -198 & -114 &
  0.5 & 0.8 & 2231($\Lambda\Lambda$) \\
    & ccs & 2059 & -318 & -172 & 1.0 & 0.6 & 2055 & -321 & -176 &
  1.0 & 0.6 &  \\
    & scb & 2328 & -194 & -58 & 1.0 & 0.6 & 2322 & -198 & -62 &
  1.0 & 0.6 &  \\
    & ccb & 2222 & -155 & -9 & 1.0 & 0.6 & 2218 & -159 & -14 &
  1.0 & 0.6 &  \\
-202 & scs & 2297 & -217 & -175 & 0.6 & 1.1 & 2307 & -215 & -165 &
  0.6 & 1.0 & 2472(N$\Xi^*$) \\
    & ccs & 2205 & -318 & -268 & 1.0 & 1.0 & 2216 & -319 & -257 &
  1.0 & 0.9 & 2397(N$\Xi\pi$) \\
    & scb & 2478 & -171 & -99 & 0.7 & 0.8 & 2476 & -173 & -102 &
  0.7 & 0.8 &  \\
    & ccb & 2369 & -182 & -103 & 1.0 & 1.8 & 2367 & -184 & -106 &
  1.0 & 0.8 &  \\
-213 & scs & 2432 & -363 & -333 & 1.0 & 1.2 & 2442 & -359 & -324 &
  1.0 & 1.1 & 2765($\Delta\Xi^*$) \\
    & ccs & 2432 & -363 & -333 & 1.0 & 1.2 & 2442 & -359 & -324 &
  1.0 & 1.1 & 2690($\Delta\Xi\pi$) \\
    & scb & 2559 & -252 & -210 & 1.0 & 1.0 & 2560 & -251 & -209 &
  1.0 & 1.0 & 2511($\Lambda\Lambda\pi\pi$) \\
    & ccb & 2556 & -252 & -209 & 1.0 & 1.0 & 2557 & -251 & -209 &
  1.0 & 1.0 &  \\
-220 & scs & 2394 & -11 & 8 & 0.1 & 1.6 & 2395 & -10 & 9 &
  0.1 & 1.6 & 2386($\Sigma\Sigma$) \\
    & ccs & 2394 & -11 & 8 & 0.1 & 1.6 & 2395 & -10 & 9 &
  0.1 & 1.6 &  \\
    & scb & 2393 & -12 & 7 & 0.1 & 1.6 & 2397 & -11 & 11 &
  0.1 & 1.5 &  \\
    & ccb & 2393 & -12 & 7 & 0.1 & 1.6 & 2397 & -11 & 11 &
  0.1 & 1.5 &  \\  \hline
-3$\frac{3}{2}$3 & scs & 2570 & -363 & -335 & 1.0 & 1.2 & 2579 & -359 & -325
  & 1.0 & 1.1 & 2904($\Delta\Omega$) \\
    & ccs & 2570 & -363 & -335 & 1.0 & 1.2 & 2579 & -359 & -325
  & 1.0 & 1.1 & 2788($\Lambda\Xi^*\pi$) \\
    & scb & 2767 & -201 & -151 & 1.0 & 0.9 & 2766 & -202 & -152
  & 1.0 & 0.9 & 2714($\Lambda\Xi\pi\pi$) \\
    & ccb & 2754 & -201 & -150 & 1.0 & 0.9 & 2754 & -201 & -150
  & 1.0 & 0.9 &  \\
-3$\frac{3}{2}$1 & scs & 2510 & -22 & -1 & 0.1 & 1.5 & 2511 & -20 & 0
 & 0.1 & 1.5 & 2511($\Sigma\Xi$) \\
  & ccs & 2512 & -23 & -0 & 0.2 & 1.4 & 2518 & -21 & 7
  & 0.2 & 1.3 &  \\
  & scb & 2525 & -25 & 14 & 0.2 & 1.1 & 2525 & -25 & 14
  & 0.2 & 1.1 &  \\
  & ccb & 2525 & -25 & 14 & 0.2 & 1.1 & 2525 & -25 & 14
  & 0.2 & 1.1 &  \\
-3$\frac{1}{2}$2 & scs & 2394 & -266 & -218 & 0.8 & 1.0 & 2394 & -266 & -217
  & 0.8 & 1.0 & 2611(N$\Omega$) \\
  & ccs & 2342 & -318 & -269 & 1.0 & 1.0 & 2353 & -319 & -259
  & 1.0 & 0.9 & 2574($\Lambda\Xi\pi$) \\
  & scb & 2556 & -267 & -147 & 1.0 & 0.6 & 2552 & -271 & -151
  & 1.0 & 0.6 &  \\
  & ccb & 2552 & -195 & -60 & 1.0 & 0.6 & 2548 & -198 & -64
  & 1.0 & 0.6 &  \\
-3$\frac{1}{2}$1 & scs & 2394 & -266 & -218 & 0.8 & 1.0 & 2394 & -266 & -217
  & 0.8 & 1.0 & 2434($\Lambda\Xi$) \\
  & ccs & 2575 & -107 & -59 & 0.4 & 1.0 & 2385 & -108 & -48
  & 0.5 & 0.9 &  \\
  & scb & 2543 & -84 & 109 & 1.0 & 0.5 & 2540 & -87 & 106
  & 1.0 & 0.5 &  \\
  & ccb & 2527 & -100 & 93 & 1.0 & 0.5 & 2523 & -104 & 89
  & 1.0 & 0.5 &  \\
  \hline
-400 & scs & 2632 & -24 & -4 & 0.1 & 1.5  & 2633 & -23 & -3
  & 0.1 & 1.5 & 2636($\Xi\Xi$) \\
  & ccs & 2632 & -24 & -4 & 0.1 & 1.5 & 2633 & -23 & -3
  & 0.1 & 1.5 &  \\
  & scb & 2643 & -24 & 7 & 0.2 & 1.2 & 2643 & -24 & 7
  & 0.1 & 1.2 &  \\
  & ccb & 2643 & -24 & 7 & 0.2 & 1.2 & 2643 & -24 & 7
  & 0.1 & 1.2 &  \\ \hline
-600 & scs & 3287 & -74 & -58 & 0.2 & 1.5 & 3291 & -69 & -54
  & 0.2 & 1.5 & 3345($\Omega\Omega$) \\
  & scb & 3351 & -29 & -6 & 0.1 & 1.0 & 3350 & -30 & 5
  & 0.1 & 1.0 &  \\ \hline
\end{tabular}
}
\end{center}
\pagebreak
\vspace*{-1.15in}
\begin{center}
Table IIb.  $\nu = 0.46 fm^{-2}$\\
{\footnotesize

\begin{tabular}{|c|c|cllcc||l|} \hline
   &   & \multicolumn{5}{c||}{(6B)} & Threshold
  \\ \hline
001 & scs & 1885 & -30 & 7 & 0.2 & 1.3 & 1878(NN) \\
    & ccs & 1878 & -31 & 0 & 0.2 & 1.4 & \\
010 & scs & 1888 & -14 & 10 & 0.1 & 1.6 & 1878(NN) \\
    & ccs & 1888 & -14 & 10 & 0.1 & 1.6 & \\
003 & scs & 2112 & -385 & -352 & 1.0 & 1.2 & 2464($\Delta\Delta$) \\
    &  &  &  &  &  &  & 2158(NN$\pi\pi$)  \\ \hline
-1$\frac{1}{2}$3 & scs & 2263 & -385 & -353 & 1.0 & 1.2 &
2617($\Delta\Sigma^*$) \\
     & ccs & 2263 & -385 & -353 & 1.0 & 1.2 & 2335(N$\Lambda\pi\pi$) \\
    & scb & 2322 & -331 & -294 & 1.0 & 1.1 & \\
    & ccb & 2322 & -331 & -294 & 1.0 & 1.1 & \\
1$\frac{3}{2}$0 & scs & 2132 & -29 & 0 & 0.2 & 1.4 & 2132(N$\Sigma$) \\
    & ccs & 2132 & -29 & 0 & 0.2 & 1.4 & \\
    & scb & 2134 & -23 & 2 & 0.1 & 1.5 & \\
    & ccb & 2134 & -23 & 2 & 0.1 & 1.5 & \\
  \hline
-200 & scs & 2128 & -212 & -129 & 0.6 & 0.8 & 2231($\Lambda\Lambda$) \\
    & ccs & 2043 & -333 & -188 & 1.0 & 0.6 & \\
    & scb & 2312 & -210 & -74 & 1.0 & 0.6 & \\
    & ccb & 2206 & -170 & -25 & 1.0 & 0.6 & \\
-202 & scs & 2278 & -236 & -195 & 0.6 & 1.1 & 2472(N$\Xi^*$) \\
    & ccs & 2185 & -337 & -287 & 1.0 & 1.0 & 2397(N$\Xi\pi$) \\
    & scb & 2461 & -188 & -117 & 1.0 & 0.8 & \\
    & ccb & 2352 & -199 & -121 & 1.0 & 0.8 & \\
-213 & scs & 2410 & -385 & -355 & 1.0 & 1.2 & 2765($\Delta\Xi^*$) \\
    & ccs & 2410 & -385 & -355 & 1.0 & 1.2 & 2690($\Delta\Xi\pi$) \\
    & scb & 2540 & -271 & -229 & 1.0 & 1.0 & 2511($\Lambda\Lambda\pi\pi$) \\
    & ccb & 2537 & -271 & -229 & 1.0 & 1.0 & \\
-220 & scs & 2391 & -14 & 5 & 0.1 & 1.6 & 2386($\Sigma\Sigma$) \\
    & ccs & 2391 & -14 & 5 & 0.1 & 1.6 & \\
    & scb & 2393 & -18 & 7 & 0.2 & 1.4 & \\
    & ccb & 2393 & -18 & 7 & 0.2 & 1.4 & \\  \hline
-3$\frac{3}{2}$3 & scs & 2548 & -385 & -356 & 1.0 & 1.2 & 2904($\Delta\Omega$)
\\
    & ccs & 2548 & -385 & -356 & 1.0 & 1.2 & 2788($\Lambda\Xi^*\pi$) \\
    & scb & 2749 & -219 & -169 & 1.0 & 0.9 & 2714($\Lambda\Xi\pi\pi$) \\
    & ccb & 2736 & -219 & -168 & 1.0 & 0.9 & \\
-3$\frac{3}{2}$1 & scs & 2509 & -30 & -2 & 0.2 & 1.3 & 2511($\Sigma\Xi$) \\
  & ccs & 2504 & -31 & -8 & 0.2 & 1.4 & \\
  & scb & 2515 & -35 & 3 & 0.3 & 1.1 & \\
  & ccb & 2514 & -34 & 2 & 0.3 & 1.1 & \\
-3$\frac{1}{2}$2 & scs & 2375 & -285 & -237 & 0.9 & 1.0 & 2611(N$\Omega$) \\
  & ccs & 2323 & -337 & -289 & 1.0 & 1.0 & 2574($\Lambda\Xi\pi$) \\
  & scb & 2540 & -283 & -163 & 1.0 & 0.6 & \\
  & ccb & 2536 & -210 & -76 & 1.0 & 0.6 & \\
-3$\frac{1}{2}$1 & scs & 2375 & -285 & -237 & 0.9 & 1.0 & 2434($\Lambda\Xi$) \\
  & ccs & 2358 & -124 & -75 & 0.5 & 1.0 & \\
  & scb & 2469 & -99 & 36 & 1.0 & 0.6 & \\
  & ccb & 2452 & -116 & 19 & 1.0 & 0.6 & \\
  \hline
-400 & scs & 2630 & -29 & -6 & 0.2 & 1.4 & 2636($\Xi\Xi$) \\
  & ccs & 2630 & -29 & -6 & 0.2 & 1.4 & \\
  & scb & 2642 & -31 & 6 & 0.2 & 1.1 & \\
  & ccb & 2641 & -32 & 5 & 0.2 & 1.1 & \\ \hline
-600 & scs & 3274 & -86 & -71 & 0.3 & 1.5 & 3345($\Omega\Omega$) \\
  & scb & 3341 & -38 & -4 & 0.2 & 1.0 & \\ \hline
\end{tabular}
}
\end{center}
\pagebreak
\vspace*{-1.20in}
\begin{center}
Table IIc.  $\mu = 1.6 fm^{-2}, \nu = 0.6 fm^{-2}$\\
{\footnotesize

\begin{tabular}{|c|c|cllcc||cllcc|l|} \hline
   &   & \multicolumn{5}{c||}{(6B)} &  \multicolumn{5}{c|}{(6A)} & \\ \hline
$SIJ$ &  & M$_{\alpha}$ & $V_{\alpha}$ & $B_{\alpha}$ & $\epsilon$ & $s_0$ &
 M$_{\alpha}$ & $V_{\alpha}$ & $B_{\alpha}$ & $\epsilon$ & $s_0$ & Threshold
  \\ \hline
001 & scs & 1869 & -46 & 9 & 0.2 & 1.3 & 1873 & -42 & 5 & 0.2 & 1.3 &
  1878(NN) \\
    & ccs & 1867 & -48 & 11 & 0.3 & 1.3 & 1873 & -42 & 5 & 0.2 & 1.3 & \\
010 & scs & 1890 & -20 & 12 & 0.2 & 1.4 & 1885 & -18 & 7 & 0.1 & 1.6 &
  1878(NN) \\
    & ccs & 1888 & -22 & 10 & 0.2 & 1.4 & 1885 & -18 & 7 & 0.1 & 1.6 & \\
003 & scs & 2074 & -422 & -390 & 1.0 & 1.2 & 2084 & -413 & -380 & 1.0 & 1.2 &
    2464($\Delta\Delta$) \\
    &   &  & &  &  &  & &  &  &  &  & 2158(NN$\pi\pi$)  \\ \hline
-1$\frac{1}{2}$3 & scs & 2225 & -422 & -390 & 1.0 & 1.2 & 2235 & -413 & -382
 & 1.0 & 1.2 & 2617($\Delta\Sigma^*$) \\
     & ccs & 2225 & -422 & -390 & 1.0 & 1.2 & 2235 & -413 & -382 &
  1.0 & 1.2 & 2335(N$\Lambda\pi\pi$) \\
    & scb & 2285 & -369 & -332 & 1.0 & 1.1 & 2292 & -361 & -324 & 1.0 & 1.1 &
    \\
    & ccb & 2285 & -369 & -332 & 1.0 & 1.1 & 2292 & -361 & -324 & 1.0 & 1.1 &
\\
1$\frac{3}{2}$0 & scs & 2119 & -41 & 13 & 0.2 & 1.4 & 2123 & -37 & 9 &
  0.2 & 1.4 & 2132(N$\Sigma$) \\
    & ccs & 2119 & -41 & 13 & 0.2 & 1.4 & 2123 & -37 & 9 &
  0.2 & 1.4 &  \\
    & scb & 2126 & -34 & 6 & 0.2 & 1.4 & 2130 & -30 & 2 &
  0.2 & 1.4 &  \\
    & ccb & 2126 & -34 & 6 & 0.2 & 1.4 & 2130 & -30 & 2 &
  0.2 & 1.4 &  \\
  \hline
-200 & scs & 2097 & -244 & -161 & 1.0 & 0.8 & 2097 & -244 & -161 &
  0.9 & 0.8 & 2231($\Lambda\Lambda$) \\
    & ccs & 1974 & -364 & -257 & 1.0 & 0.7 & 2012 & -365 & -220 &
  1.0 & 0.6 &  \\
    & scb & 2282 & -240 & -104 & 1.0 & 0.6 & 2281 & -241 & -105 &
  1.0 & 0.6 &  \\
    & ccb & 2176 & -200 & -55 & 1.0 & 0.6 & 2175 & -202 & -56 &
  1.0 & 0.6 &  \\
-202 & scs & 2242 & -272 & -231 & 0.8 & 1.1 & 2249 & -265 & -223 &
  0.7 & 1.1 & 2472(N$\Xi^*$) \\
    & ccs & 2150 & -373 & -323 & 1.0 & 1.0 & 2255 & -368 & -318 &
  1.0 & 1.0 & 2397(N$\Xi\pi$) \\
    & scb & 2413 & -221 & -164 & 0.9 & 0.9 & 2429 & -219 & -148 &
  1.0 & 0.8 &  \\
    & ccb & 2319 & -231 & -153 & 1.0 & 0.8 & 2320 & -230 & -152 &
  1.0 & 0.8 &  \\
-213 & scs & 2372 & -423 & -393 & 1.0 & 1.2 & 2382 & -413 & -383 &
  1.0 & 1.2 & 2765($\Delta\Xi^*$) \\
    & ccs & 2372 & -423 & -393 & 1.0 & 1.2 & 2382 & -413 & -383 &
  1.0 & 1.2 & 2690($\Delta\Xi\pi$) \\
    & scb & 2496 & -308 & -273 & 1.0 & 1.1 & 2509 & -302 & -260 &
  1.0 & 1.0 & 2511($\Lambda\Lambda\pi\pi$) \\
    & ccb & 2493 & -308 & -272 & 1.0 & 1.1 & 2506 & -302 & -259 &
  1.0 & 1.0 &  \\
-220 & scs & 2391 & -20 & 5 & 0.2 & 1.4 & 2388 & -18 & 2 &
  0.1 & 1.6 & 2386($\Sigma\Sigma$) \\
    & ccs & 2390 & -21 & 4 & 0.2 & 1.4 & 2388 & -18 & 2 &
  0.1 & 1.6 &  \\
    & scb & 2382 & -33 & -4 & 0.2 & 1.3 & 2386 & -29 & 0 &
  0.2 & 1.3 &  \\
    & ccb & 2381 & -34 & -5 & 0.2 & 1.3 & 2385 & -30 & -1 &
  0.2 & 1.3 &  \\  \hline
-3$\frac{3}{2}$3 & scs & 2510 & -423 & -394 & 1.0 & 1.2 & 2520 & -413 & -384
  & 1.0 & 1.2 & 2904($\Delta\Omega$) \\
    & ccs & 2510 & -423 & -394 & 1.0 & 1.2 & 2520 & -413 & -384
  & 1.0 & 1.2 & 2788($\Lambda\Xi^*\pi$) \\
    & scb & 2704 & -254 & -214 & 1.0 & 1.0 & 2717 & -250 & -201
  & 1.0 & 0.9 & 2714($\Lambda\Xi\pi\pi$) \\
    & ccb & 2691 & -255 & -213 & 1.0 & 1.0 & 2705 & -250 & -200
  & 1.0 & 0.9 &  \\
-3$\frac{3}{2}$1 & scs & 2493 & -46 & -18 & 0.2 & 1.3 & 2497 & -42 & 14
  & 0.2 & 1.3 & 2511($\Sigma\Xi$) \\
  & ccs & 2491 & -48 & -20 & 0.3 & 1.3 & 2497 & -42 & 14
  & 0.2 & 1.3 &  \\
  & scb & 2498 & -60 & 13 & 0.4 & 1.0 & 2501 & -57 & 10
  & 0.3 & 1.0 &  \\
  & ccb & 2497 & -61 & 14 & 0.4 & 1.0 & 2500 & -58 & 11
  & 0.4 & 1.0 &  \\
-3$\frac{1}{2}$2 & scs & 2339 & -321 & -272 & 1.0 & 1.0 & 2344 & -316 & -268
  & 1.0 & 1.0 & 2611(N$\Omega$) \\
  & ccs & 2287 & -373 & -324 & 1.0 & 1.0 & 2292 & -368 & -319
  & 1.0 & 1.0 & 2574($\Lambda\Xi\pi$) \\
  & scb & 2510 & -313 & -193 & 1.0 & 0.6 & 2508 & -304 & -194
  & 1.0 & 0.6 &  \\
  & ccb & 2506& -240 & -106 & 1.0 & 0.6 & 2504 & -242 & -107
  & 1.0 & 0.6 &  \\
-3$\frac{1}{2}$1 & scs & 2339 & -321 & -272 & 1.0 & 1.0 & 2344 & -316 & -268
  & 1.0 & 1.0 & 2434($\Lambda\Xi$) \\
  & ccs & 2337 & -156 & -97 & 1.0 & 0.9 & 2340 & -154 & -94
  & 1.0 & 0.9 &  \\
  & scb & 2474 & -167 & -38 & 1.0 & 0.6 & 2437 & -131 & 3
  & 1.0 & 0.6 &  \\
  & ccb & 2422 & -146 & 11 & 1.0 & 0.6 & 2421 & -147 & 13
  & 1.0 & 0.6 &  \\
  \hline
-400 & scs & 2617 & -41 & -19 & 0.2 & 1.4 & 2622 & -37 & -15
  & 0.2 & 1.4 & 2636($\Xi\Xi$) \\
  & ccs & 2617 & -41 & -19 & 0.2 & 1.4 & 2622 & -37 & -15
  & 0.2 & 1.4 &  \\
  & scb & 2628 & -45 & -8 & 0.3 & 1.1 & 2630 & -43 & -6
  & 0.2 & 1.1 &  \\
  & ccb & 2627 & -46 & -9 & 0.3 & 1.1 & 2630 & -43 & -6
  & 0.2 & 1.1 &  \\ \hline
-600 & scs & 3251 & -112 & -94 & 0.3 & 1.4 & 3260 & -102 & -84
  & 0.3 & 1.4 & 3345($\Omega\Omega$) \\
  & scb & 3376 & -65 & 32 & 1.0 & 0.6 & 3375 & -67 & 30
  & 1.0 & 0.6 &  \\ \hline
\end{tabular}
}
\end{center}
\end{document}